\documentclass{icrc2009}

\usepackage{graphicx}   
\usepackage{caption}   
\usepackage{fixltx2e}
\usepackage{url}

\newcommand{\shorttitle}[1]%
{\markboth{Proceedings of the 31\MakeLowercase{$^{st}$} ICRC, {\L}\'{o}d\'{z} 2009}{#1} }
\newcommand{\etal}{\MakeLowercase{\textit{et al. }}} 

\begin{document}
\title{Energy Determination of Solar Neutrons by the SEDA-AP 
                    on-board JEM of ISS}

\author{\IEEEauthorblockN{K. Koga, T. Goka, H. Matsumoto and T. Obara\IEEEauthorrefmark{1},Y. Muraki and T. Yamamoto\IEEEauthorrefmark{2}}
        
\IEEEauthorblockA{\IEEEauthorrefmark{1} Space Environment group, JAXA, Tsukuba, 305-8505, Japan}
\IEEEauthorblockA{\IEEEauthorrefmark{2}Department of Physics, Faculty of Science, Konan University, Kobe 658-8501, Japan}}

\shorttitle{Muraki \etal solar neutron telescope onboard ISS}
\maketitle

\begin{abstract}
A solar neutron detector will be launched on the Japan Exposure Module (JEM) of the International Space Station (ISS) in  June 2009. The detector comprises scintillation fiber and is designated as the Space Environment Data Acquisition (SEDA)-FIB. It tracks the recoil protons induced by neutrons and measurement of the proton energy using the range method.   The energy resolution of the detector was measured using the proton beam at Riken.   Herein, we report the energy resolution of the FIB detector by two different methods.

\end{abstract}

 
\section{Introduction}
 Measurement of the radiation effects imparted to astronauts on the space station is an important task for the International Space Station (ISS) mission.  Numerous high-energy protons are produced at the solar surface when a huge solar flare occurs.   They arrive at the ISS a few hours later. These protons pose a strong radiation hazard not only to the astronauts but also to the electronic devices on-board the ISS.   An early prediction of the arrival of the Solar Energetic Particles (SEPs) is expected to minimize such radiation hazards.  For realization of such a forecast, a new type of solar neutron detector is proposed for mounting on the ISS.  In April 1979, the plan was accepted by the Space Development Committee of the Ministry of Science of Japan as a JEM payload \cite{Minute}.
 
The new neutron detector can measure the neutron energy; it can also detect the direction of the arrival of neutrons.  Those functions are very important not only to identify when solar neutrons depart from the Sun but also to assess radiation hazards for human bodies and radiation damage to electronics.  These new functions will enable more precise understanding not only of the production mechanisms of the Solar Energetic Particles (SEPs) at the Sun surface, but also the radiation effects imparted by them.  We can determine whether they are produced indirectly by the wall of ISS or directly by solar neutrons.

\section{New Solar Neutron Detector}
The new neutron detector can measure the energy of solar neutrons and the Albedo neutrons from the Earth.   The detector comprises fine blocks of the scintillation plates; the size of one  scintillator plate is 6mm$\times$3mm$\times$96mm.  The neutron tracks are identifiable as the tracks of protons that are produced through n-p interaction processes in the scintillator plates.  The arrival direction and the track length of protons are detected by two multi-anode photomultipliers (H4140-20; Hamamatsu Photonics K.K.) looking perpendicularly from the X direction and Y direction (Fig. 1).   Thereby, the detector can identify the arrival direction of neutrons with accuracy of $\pm$(1.8$-$45) deg, depending on the track length.   Consequently, we can identify whether those neutrons have come from the Sun or the Earth or from the wall of the ISS because the direction of the Sun is measured using a position sensor on-board the ISS. 
The neutron's| energy can be estimated by measuring the range of protons.  The neutron detector designated as FIB in SEDA can measure the energy of neutrons between 30 MeV and 100 MeV.   

However we noticed another method to obtain the energy of protons.  The intensity of photons deposited into each block of the scintillator plate will be measured one-by-one by the Analog-Digital-Converter.  Therefore the sum of all ADC values deposited on the scintillator plate along a track must be proportional to the kinetic energy of protons.   Herein we call this method "total photon counting method".   The deposited energy in each thin plastic plate will be measured by the 256 channel multi-anode photomultipliers one by one, switching the ADC electronically.   The diagram of the electric logic is shown in Figure 2.    In this conference, we will report the energy resolution of the total photon counting method in oral presentation, comparing them with that of the range method.

Simultaneously, the Bonner Ball Detector (BBD) will measure neutrons with energies in the range of (0.03 eV$-$15 MeV) in SEDA.   The BBD detector cannot identify the arrival direction of neutrons|.  The operational principle of the BBD is the same as the neutron monitors that are used throughout the world to measure the long-term modulation of cosmic ray intensity, but the detector is small.
The FIB and the BBD detectors are mounted in a large box called the Space Environment Data Acquisition equipment-Attached Payload (SEDA-AP), which weighs 450 kg.  It will be launched on the Japan Exposure Module (JEM) of the ISS on 13 June 2009 by the Endeavour space shuttle.  Details of the FIB detector on-board the SEDA-AP are available elsewhere \cite{Koga, Matsumoto, Imaida, Muraki, webJAXA}.

\section{Energy Resolution of FIB detector}
The energy resolution of the FIB detector was obtained using the proton beam at Riken.  A 160 MeV proton beam was bombarded in front of the FIB.  Different proton energy was realized to install aluminum plates of various thicknesses.   The energy radiated on the FIB detector is respectively equivalent to the energy of 27 MeV, 44 MeV, 68 MeV, and 102 MeV.  The range of tracks was scanned visually to obtain the range distribution.   We have inferred that the mean value of the range corresponds to the incident proton energy, but the distribution from the mean value corresponds to the detector|s energy resolution. Consequently, the energy resolution was obtained.   The results are presented in Fig. 3.  Fitting the data to a function of 1/E, we obtained the energy resolution of the FIB to the proton tracks as $\Delta$E/E = 10$\%$/(E/50 MeV).

However, when the number of trigger events for neutrons increases to more than 16 Hz, the FIB detector cannot record the pattern of the event because the transmission rate is limited by the baud rate of the communication between the ISS and the ground-based station.   In this case, they can measure the number of layers, i.e., the range of charged particles.   Furthermore, in case the trigger rate exceeds 64 Hz, the total deposited energy will be measured merely using use of the mesh type dynode of the multi-anode photomultipliers.  At this time, the energy resolution is not as good as that of the range method: it is $\Delta$E/E = 40$\%$, being independent from energy.

\section{ Expected trigger rate and Large solar flares}
  The FIB will record various data.   The main sensor is cubic: 10cm$\times$10cm$\times$10cm.  The main sensor is covered by six plates of the plastic scintillator, which have a role of the anti-counter.  Therefore, when high-energy neutrons enter into the sensor and produce high-energy protons, the high-energy protons penetrate the main sensor and arrive at the anti-counter.   Using the anti-counter, such high-energy neutrons are not recorded.   We can remove the bottom side anti-counter using communications from the ground, but it depends on the trigger rate by the backside particles.  Detection of high-energy solar neutrons with energies $>$100 MeV is made using ground-based neutron detectors.
  
The onboard detectors of the ISS can record solar neutrons with energy of less than 100 MeV.   The delay time of solar neutrons with energy of 100 MeV is 11 min later than the light if solar neutrons are produced instantaneously at the Sun by the solar flare.   According to our estimation \cite{Imaida}, the expected flux of solar neutrons for gigantic solar flares, e.g. the 4 June 1991 event, is 15 Hz for 100 MeV.  Here we have estimated the detection efficiency of neutrons by the FIB as 10$\%$.   Therefore, we will be able to record the pattern of events with energy ranges of between 40$-$100 MeV using the FIB detector for smaller solar flares such as X=1.  This limitation is imposed merely by the available memory on the FIB.

The FIB detector can only record the total energy of the events if the trigger rate becomes greater than 64 Hz.  Such a situation is expected to occur for neutrons with energy of less than 60 MeV for the largest solar flares.  The expected background level is 1.5 Hz above the middle latitude and 0.2 Hz above the equator.  They are produced by albedo neutrons, so-called cosmic ray albedo neutrons, induced by collisions of the galactic cosmic rays with the nucleus of the upper atmosphere.

Finally in Figure 4, we present a photograph of the end-to-end test experiment that was
made at the Tsukuba space center of JAXA in the end of 2007.

\section{Acknowledgements}
The authors thank Mr. N. Mochizuki for Flight Module data analysis.

\begin{figure}
\centering
\includegraphics[width=2.7in]{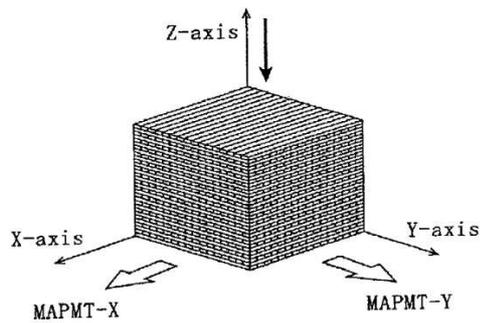}
\caption{Schematic view of the neutron detector.   It comprises scintillation plates of
3mm$\times$6mm$\times$96mm.   The signal of each plate is read out by the
multi-anode photomultipliers.  The track pattern can be recognized using 512 data points.
The total weight of the FIB detector is about 20kg, except the interface part.  }
\label{fig1}
\end{figure}

\begin{figure*}[th]
\centering
\includegraphics[width=2.5in]{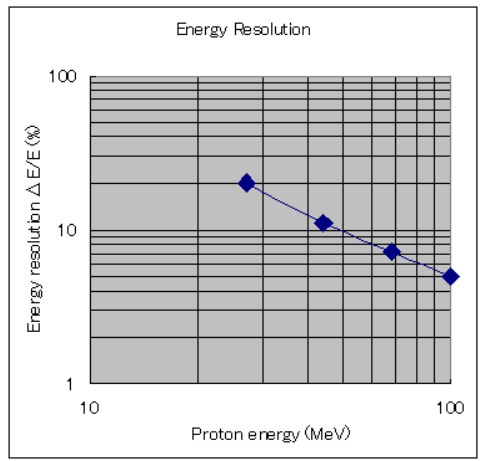}
\caption{The energy resolution of the neutron detector obtained using 
the range method is shown as a function of the energy of incoming protons.  
The energy resolution can be well fitted to a function of  $\Delta$E/E=10$\%$/(E/50MeV).}
\label{fig3}
\end{figure*}

 
\begin{figure*}[th]
 \centering
 \includegraphics[width=6.0in]{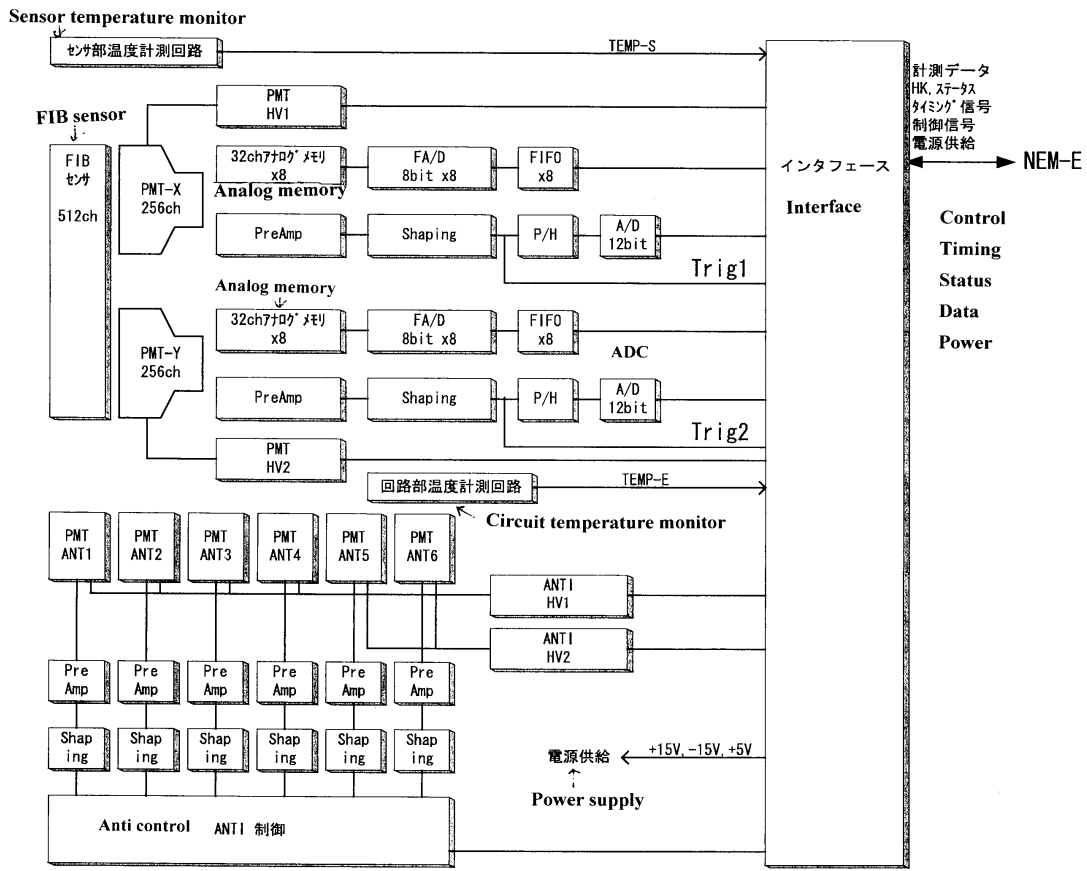}
 \caption{The block diagram of the data acquisition.   The signals will be detected by the 
two multi-anode photomultipliers labeled PMT-X and PMT-Y.   The pulse height of those
signals will be counted by the A/D converter after peak holding.    The analog signal
will be also stored in the Analog memory and via Flash ADC, each charge will be
counted and recorded.   Six photomultipliers  are prepared to reject the charged particles 
coming from each side of the central cubic neutron sensor.   The power supply and the
temperature monitors of the circuits and sensors are also prepared.   Those signals are 
sent to the common bus line of the ISS and will be down-loaded to the Tsukuba data center.}
\label{fig2}
\end{figure*}


\begin{figure*}[th]
\centering
 \includegraphics[width=6in]{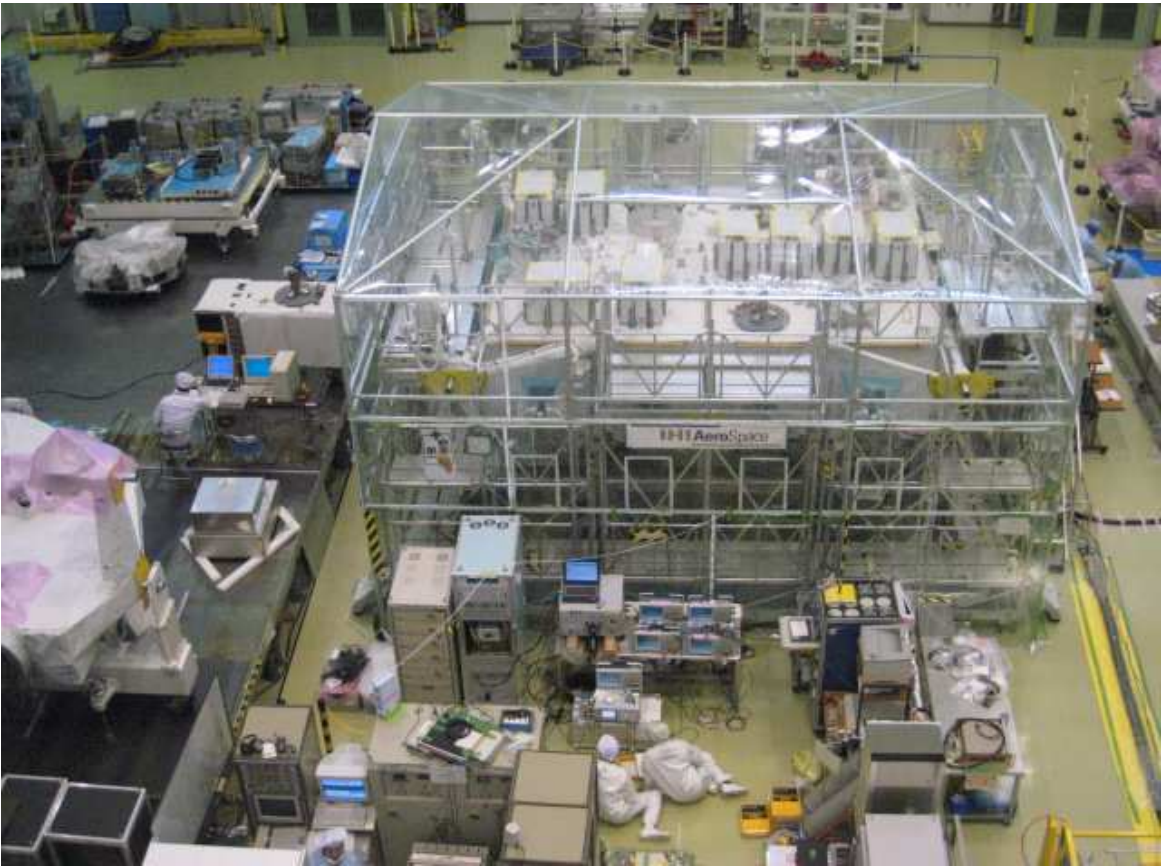}
 \caption{A photograph of the end-to-end test experiment.   The photograph was 
taken in JAXA Tsukuba space center on December 17, 2007.  The photograph  was taken by
one of the authors.  The test experiment was made between the SEDA-AP 
(left side out of the tent) and the Kibo exposed facility (inside the tent).}
\label{fig4}
\end{figure*}

\end{document}